**New Symmetries in Crystals and Handed Structures**

Venkatraman Gopalan[1] and Daniel B. Litvin[2]

[1]*Department of Materials Science and Engineering, Pennsylvania State University, University Park, PA, 16803*

[2] *Department of Physics, Eberly College of Science, The Pennsylvania State University, Penn State Berks, P.O. Box 7009, Reading, PA 19610*

**Symmetry is a powerful framework to perceive and predict the physical world. The structure of materials is described by a combination of rotations, rotation-inversions and translational symmetries[1,2]. By recognizing the reversal of static structural rotations between clockwise and counterclockwise directions as a distinct symmetry operation, here we show that there are many more structural symmetries than are currently recognized in right- or left-handed handed helices[3], spirals[4], and in antidistorted structures[5] composed equally of rotations of both handedness. For example, though a helix or spiral cannot possess conventional mirror or inversion symmetries, they can possess them in combination with the rotation reversal symmetry. Similarly, we show that many antidistorted perovskites possess twice the number of symmetry elements as conventionally identified. These new symmetries predict new forms for "roto" properties that relate to static rotations, such as rotoelectricity, piezorotation, and rotomagnetism. They also enable symmetry-based search for new phenomena, such as multiferroicity involving a coupling of spins, electric polarization and static rotations. This work is relevant to structure-property relationships in all material structures with static rotations such as minerals[5,6], polymers[7], proteins[8], and engineered structures[9].**

An introduction to the fields of materials science and solid state physics typically begins with a discussion of symmetry and crystal structure, which serves as the foundation for structure-property relationships in materials[10]. The symmetry of an object is a set of operations, called a symmetry group, which brings the object into self-congruence. The spatial symmetry operations of translation, rotation, and roto-inversion, give rise to 32 crystallographic point groups and 230 crystallographic space groups[1,11]. The addition of time reversal operation 1' that inverts time and hence magnetic moments (Figure 1a, b), extends the symmetry groups to an additional 90 magnetic point groups (MPG) and 1421 magnetic space groups that describe magnetic structures[12]. Of the many such antisymmetry operations that can be defined to switch between two states of a given property of the atoms, inversion, $\overline{1}$, (that inverts spatial coordinates about a point) and time reversal 1' are of particular significance. Figure 1e shows how these symmetries transform polar (such as an electric dipole) and axial (such as a spin) vectors. While $\overline{1}$ can invert all polar vectors, and 1' can invert time dependent polar or axial vectors, neither 1' nor $\overline{1}$ can invert a time independent axial vector, such as those that describe static rotations, the subject of this letter.

Here we introduce a new antisymmetry operation, namely, rotation reversal symmetry $1^{\Phi}$, that reverses the sense of static rotations in structures such as in Fig. 1c,d. Static structural distortions are typically described by normal mode analysis[13,14,15]. We illustrate that $1^{\Phi}$ can identify new *roto symmetry groups* corresponding to primary normal modes of static rotations; these roto groups can have many more symmetry elements than the conventional symmetry group assignments. Furthermore, we show that roto symmetries can predict new forms for "roto" property tensors that relate static rotations to physical quantities such as temperature, fields, and stress. Denev et al.[16] first suggested that antidistorted perovskite structures may possess more symmetry elements than is obvious. However, they used time reversal symmetry to describe an optical property tensor that tracks a structural phase transition. Time reversal however cannot be used to describe the static rotations themselves in a structure, hence the need for a new symmetry operation introduced here.

Consider Figure 1c for example. It shows a planar antidistorted structure constructed of square motifs with atoms at each corner denoted by 1-2-3-4, in two orientations, +Φ and -Φ about a rotation axis, color coded aqua and orange respectively. The crystallographic point group symmetry of this planar array of atoms is $m_x m_y 2_z$, where subscripts indicate the orientations of the mirror ($m$) and 2-fold rotation axis. Consequently this atom arrangement is classified by the symbol:

$$(m_x m_y 2_z \mid r_1, r_2, r_3, r_4) \qquad (1)$$

That is, all atoms of this arrangement can be obtained by applying the operations of the group $m_x m_y 2_z$ to the four atoms at positions $r_1$ through $r_4$. The rotation angle of each motif can be described by the static rotational moment given by the axial vector, $\mathbf{\Phi}(r) \sim \hat{r}_i \times \hat{r}'_i$, which is a small common *static* rotation angle Φ of a set of atoms, indexed $i$ (=1-4 in Fig 1c,d), about a given axis passing through the centre of mass of the set of atoms being rotated. The vector $\mathbf{\Phi}(r)$ is thus an axial vector with magnitude Φ and a direction along the direction of the rotation axis, given by the right hand rule. We define $1^{\Phi}$, defined formally in the Methods Section, as an operation that reverses the sign of Φ→-Φ *without translating the centre of mass* of the relevant set of atoms. (In this sense, $1^{\Phi}$ is distinct from other spatial symmetries such as say a mirror or a 2-fold rotation axis that might also invert a static rotation, but will also translate the centre of mass of the rotating motifs in space). Thus $1^{\Phi}$ transforms the structure between Figs. 1c and 1d.

Then $4_z^\Phi$, a 4-fold rotation about $z$ followed by $1^\Phi$, is also a symmetry of this structure. The atom arrangement of Figure 1b is now classified as

$$(4_z^\Phi m_x m_{xy}^\Phi \mid r_1, r_2) \qquad (2)$$

where the subscript *xy* indicates the <110> directions. The group in Eq. (2) contains *twice* the number of symmetry elements as the group in Eq. (1) (8 versus 4). More information on the geometric structure of the arrangement is now stored within the symmetry group. As a consequence, additional information that is needed in the form of atomic positions, $r$ decreases from (1) to (2). Note the striking similarity to Fig. 1a, where the point group without time reversal symmetry, 1', is $m_x m_y 2_z$, and with 1' is $4'_z m_x m'_{xy}$, resulting in twice the number of symmetry elements. This is important since the extra symmetry information influences property predictions, which it otherwise would not.

To illustrate the interplay between static and dynamic rotational operations, let us consider the possibility that a motif such as in Figure 1c possesses both a static rotation, $+\Phi$, and a magnetic moment, $+M$. Time reversal, $1'$ will reverse $M$ and transform this motif to the state ($+\Phi$, $-M$), while $1^\Phi$ will reverse $\Phi$ and hence the motif to state ($-\Phi$, $+M$). Similarly, $1^{\Phi'} = 1^\Phi \cdot 1'$ will reverse both $\Phi$ and $M$ to transform the motif to state ($-\Phi$, $-M$).

We now generalize the classical 32 point groups and 230 space groups by combining the elements of these groups with the operations 1' and $1^\Phi$. The resulting new 624 roto-point groups and 17,807 roto-space groups are isomorphic to the double antisymmetry point groups and space groups[17,18]. In Figure 2a, we have subdivided these groups into eight subtypes according to whether or not the groups contain as elements each of the three operations 1', $1^\Phi$, and $\bar{1}$. In Figure 2b, we list the symmetry space groups of arrangement of spins, static rotations, or electric dipoles, or combinations thereof. In the Supplementary Table 1, we list the point groups of space groups[19] that are invariance groups of a non-zero spin, static rotation, or electric dipole moment, or combinations thereof. Note that there are only 12 multiferroic groups possessing all three properties, which can allow, for example, a symmetry-based search for coupling between polarization, spins and static rotations in materials. Note also that many of these 12 groups are simple rotational point groups such as 1, 2, 3, 4, 6, which suggests that crystals that contains only simple rotation axes or screw axes, such a protein crystals, are potential candidates.

We next discuss octahedral ($X_6$) rotations (also called antidistortive tilts), which are the most common phase transitions in perovskite structures, $ABX_3$, where A, B are cations and X an anion. An example of a non-magnetic antidistorted structure in a cubic perovskite structure is shown in Figures 3a, conventionally described by Glazer notation (Methods summary) as $a_o^+a_o^+c_o^+$ [5], and by the orthorhombic space group, $Immm1'$. However, we identify the roto space group for this structure as the tetragonal R-group, $I4^{\Phi}/mmm^{\Phi}1'$ (Figure 3b), which has twice the number of elements as compared with $Immm1'$. Supplementary Table 2 lists the original 23 cubic non-polar Glazer groups and their newly derived R- group symmetries. Note for example that $Immm1'$ applies to many structures, such as $a_o^+b_o^+c_o^+$, $a_o^ob_o^+c_o^+$, and $a_o^+a_o^+c_o^+$ while the roto symmetry group classification clearly distinguishes these three. Figure 3c depicts an example of a MR-group comprised of a magnetic perovskite structure with $a_o^+a_o^+c_o^+$ distortions. The conventional space group description is the magnetic group, $Im'm'm$, but the roto-group assignment is $I4^{\Phi}/mm'm^{\Phi'}$. Finally, in supplementary discussion 1, we discuss the normal modes of $NaNbO_3$, and show that the primary modes that reflect the *true* symmetry of the crystal have a combined roto symmetry group of $C_pm^{\Phi}cm1'$, which has twice the number of symmetry elements as the conventional group $Pbcm1'$ currently assigned to this structure. Similarly, the static distortions in polar bismuth ferrite, $BiFeO_3$, ignoring spins, has an RP symmetry group of $R_R3m^{\Phi}1'$, ignoring spins, and an MP symmetry (ignoring rotations, but allowing canting) of *Cc* or *Cc′*. The composite MRP-group symmetry (RP∩MP) is *Cc* or *Cc′*.

Supplementary Table 3 classifies a range of roto properties based on symmetry. The first named "roto" property in literature was rotostriction by Haun[20]. The form of these tensors and others can easily be derived using the coordinate transformation rules (Method's section) as well as space group approach (Supplementary discussion 3). The most commonly reported roto properties are those that are invariant to $1^{\Phi}$, supplementary discussion 2, include rotostriction[20], quadratic[21] and biquadratic[20] rotoelectricity, and rotomagnetoelectricity[22], to name a few. The properties that are more rare in literature are those that invert under $1^{\Phi}$, supplementary discussion 2, include linear rotoelectricity[23], piezorotation[24], chiral sum frequency generation[25], torroidal magnetism[26], and linear rotomagnetism[27].

The new symmetries often predict a "roto" tensor different from that predicted by conventional symmetry. Consider the piezorotation tensor (supplementary Table 3), or the quadratic electrorotation ($\Phi_i = Q_{ijk} E_j E_k$) or the quadratic magnetorotation ($\Phi_i = Q_{ijk} B_j B_k$) tensors, all of which have the same forms. Since $\Phi_i$ is a local order parameter, using a space group analysis (see supplementary discussion 3), these property tensors are predicted by the conventional $Immm1'$ group for the structure in Fig. 3a,b to have three non-zero independent coefficients, $Q_{14} \neq Q_{25} \neq Q_{36}$. However, the correct tetragonal space group of $I4^{\Phi}/mmm^{\Phi}1'$ predicts only one non-zero independent coefficient, $Q_{14}=-Q_{25}$, while $Q_{36}=0$. Similarly, when considering the MR group in Figs. 3c,d for example, the conventional global point group of $m'm'm$ predicts a linear rotomagnetic tensor with two non-zero independent coefficients, $Q_{12} \neq Q_{21}$. In contrast, the complete MR-point group assignment of $4^{\Phi}/mm'm^{\Phi'}$ predicts only one, namely, $Q_{12}=Q_{21}$.

Finally we show that the symmetry framework presented here is applicable to any structure that possesses handedness, including helices, spirals, and crystals composed of them as motifs (e.g. protein crystals). An appropriate function $\Phi(\boldsymbol{r})$ or $\boldsymbol{r}(\Phi)$, Supplementary equations 4 and 5, is identified for each structure on which $1^{\Phi}$ can operate and reverse its handedness *without translating its center of mass in space*. (Note that though a mirror or inversion operation may also reverse the handedness of a *single* helix or spiral, in a crystal composed of many helices (such as protein crystals), they will also translate them, supplementary discussion 5.) A helix has no conventional mirror or inversion symmetry. However, Fig. 4a, a left-handed helix is related to a right-handed helix by the operation $1^{\Phi}$. A continuous *infinite* single helix (Fig. 4b) and a double helix, (Fig. 4c, supplementary discussion 6), have a point group of $(\infty/m^{\Phi})m^{\Phi}2$ (new) versus $\infty 2$ (conventional). Note that $\bar{1}^{\Phi}$ is a symmetry element of these helices. If a charge current is flowing through a helix (Fig. 4a), the point group symmetry is $(\infty/m^{\Phi'})m^{\Phi'}2'$. A continuous helix of *finite* length with integral windings (Fig. 4d) has point group symmetry of $m^{\Phi}m^{\Phi}2$ (new) versus 2(conventional). A planar spiral, Fig. 4e, has point group symmetry of $mm^{\Phi}2^{\Phi}$(new) versus $m$ (conventional). A left- and a right-handed screw $3_1$ (Fig. 4f,g and supplementary discussion 5) are related by $1^{\Phi}$, and have a point group of $(\bar{6}^{\Phi}/m^{\Phi})m^{\Phi}2$ (new) instead of 32 (conventional). Again, there are property consequences of these new symmetries; for example, the piezorotation tensor (supplementary Table 3) for the point groups 32 and $(\bar{6}^{\Phi}/m^{\Phi})m^{\Phi}2$ are different.

In conclusion, the introduction of the rotation reversal symmetry operation, $1^{\Phi}$, completes the set of antisymmetry operations along with inversion, $\bar{1}$ and time reversal, $1'$ that allow a reversal of *all* vector quantities, both polar and axial, as well as *static* and dynamic. This leads to the discovery of new symmetries in structures with static rotations, and in turn leads to new *roto* point and space group symmetries of well-known structures such as helices, spirals, and antidistorted crystals, introducing more symmetry than conventional symmetry assignments. A direct consequence of these new symmetries is their applicability to predicting the form of a wide range of roto properties (Supplementary Table 3) that relate to static rotations in materials. For example, in search of *linear rotoelectric effect* that would allow the control of the magnitude and direction of static rotations with electric field, only two (#17 and #18 in Supplementary Table 2) of the original 23 Glazer groups need be explored. In search of multiferroics with coupling between spins, polarization and static rotations, only 12 of the 624 roto point groups need be explored (Supplementary Table 1). Further, we often predict significantly different forms for these properties than is predicted from conventional symmetry identifications. Since $1^{\Phi}$ reverses the sign of a mathematical cross-product between two vectors, which is very common in describing physical quantities such as in the use of area or curl of a vector field, it is applicable to a large number of physical properties. Since static rotations are abundant in physical[5,6,10,12,22-27] chemical[3,7,25], biological[4,8] and engineering[4,9] disciplines, we expect that the new symmetries discovered in this Letter will have broad interdisciplinary relevance.

## METHODS SUMMARY

A direction of the cross product of two vectors **A** and **B,** given as $\mathbf{A} \times \mathbf{B} = AB\sin\Phi\hat{n}$, is proportional to the sign of the angle **Φ** between them, and hence will reverse sign under $1^{\Phi}$ operation. Glazer notation $a_o^+ a_o^+ c_o^-$, e.g. indicates equal magnitude of rotation of the octahedra about [100] and [010] and a different rotation about the [001] *z*-axis. A negative versus positive superscript indicates that the adjacent octahedral rotations alternate in sign or have the same sign, respectively, along that axis. The subscripts denote zero (o), positive (+) or negative (-) polar displacements along that axis. The MR group in Figure 3b for example, arises as the intersection (R-group∩M-Group) of the symmetries of the R-group sublattice ($I4^{\Phi}/mmm^{\Phi}1'$) and the M-group sublattice ($P4/mm'm'1^{\Phi}$). The rotational sublattice ignores the magnetic structure, and a magnetic sublattice ignores the rotational structure.

## METHODS

### Generalization of Space Group Symmetry with $1^{\Phi}$

We consider a crystal made up of identical subunits called molecules that exist in two equally populated orientations within the crystal. The atoms composing the molecule will be considered as discrete points in space and consequently a mathematical model of the crystal can be given by a scalar density function ρ(**r**) given by

$$\rho(\boldsymbol{r}) = \sum_j \sum_k \delta(\boldsymbol{r} - \boldsymbol{r}_j^{cm} - \boldsymbol{r}_{jk(n)}) \qquad (3)$$

where $\boldsymbol{r}_j^{cm}$ is the center of mass position of the $j^{th}$ molecule and $\boldsymbol{r}_{jk(n)}$ is the position vector of the $k^{th}$ atom in the $j^{th}$ molecule relative to the center of mass at $\boldsymbol{r}_j^{cm}$, and $n$ = +1, -1 represents the two orientations of the $j^{th}$ molecule.

An element of a space group **G** is denoted by $\mathrm{G} = (R\,|\,\mathbf{v}(R) + \mathbf{t})$ where $R$ is a symmetry operation, $\mathbf{v}(R)$ the non-primitive translation associated with $R$, and $\mathbf{t}$ a primitive translation. It follows that

$$(R\,|\,\mathbf{v}(R) + \mathbf{t})\rho(\boldsymbol{r}) = \sum_j \sum_k \delta(\boldsymbol{r} - R\boldsymbol{r}_j^{cm} + \mathbf{v}(R) + \mathbf{t} - R\boldsymbol{r}_{jk(n)}) \qquad (4)$$

The atom arrangement is said to be invariant under the space group element $(R\,|\,\mathbf{v}(R) + \mathbf{t})$, and the space group element is said to be a symmetry element of the crystal if $(R\,|\,\mathbf{v}(R) + \mathbf{t})\rho(\boldsymbol{r}) = \rho(\boldsymbol{r})$. The set of all such symmetry elements constitutes the space group of the crystal. We introduce a switching operation, denoted by $1^{\Phi}$ which acts to switch the orientation of the molecules:

$$1^{\Phi}\rho(\boldsymbol{r}) = \sum_j \sum_k \delta(\boldsymbol{r} - \boldsymbol{r}_j^{cm} - \boldsymbol{r}_{jk(1^{\Phi} n)}) \qquad (5)$$

$1^{\Phi}$ switches the orientation of the molecules, i.e. $1^{\Phi} n = -n$. It follows then that $(1^{\Phi})^2 = 1$ is an identity operation. With this switching operation we generalize the space group symmetry of molecular crystals: We define two types of operations: $R$ and $R^{\Phi} = R.1^{\Phi}$. The set of operations $(R \mid \mathbf{v}(R)+\mathbf{t}) = (1 \,\|\, \mathrm{G})$ and $(R^{\Phi} \mid \mathbf{v}(R)+\mathbf{t}) = (1^{\Phi} \,\|\, \mathrm{G})$ which leave the molecular crystal density function $\rho(\boldsymbol{r})$ invariant constitute the generalized space group symmetry of the molecular crystal.

These generalized space groups, where $1^{\Phi}$ commutes with $R$, have two intrinsic properties: (1) Half the elements are coupled with the switching operation and half are not, and (2) These generalized groups have the same abstract mathematical structure as, and are isomorphic to magnetic groups.

**Coordinate transformation rules for "roto" property tensors**

The linear orthogonal transformation rules for the vectors components $P_i$, $\Phi_i$ and $M_i$ from old (superscript $o$) to new (superscript $n$) coordinates by a point group symmetry operation (with matrix elements $a_{ij}$), is defined as $P_i^n = a_{ij} P_j^o$, $\Phi_i^n = (\pm)^{\Phi} (\pm)^{\Phi'} |a| a_{ij} \Phi_j^o$, and $M_i^n = (\pm)'(\pm)^{\Phi'} |a| a_{ij} M_j^o$, respectively. The $(\pm)^{\Phi}$ for example, is equal to -1 when the $1^{\Phi}$operation is associated with the symmetry operation, and is +1 otherwise. The transformation rules for roto property tensors $Q_{ijkl..}$ in Table 1 are as follows:

For row 1, Table 1: $Q^n_{ijkl..} = (\pm)^{\Phi} (\pm)^{\Phi'} |a| a_{ip} a_{jq} a_{kr} a_{ls} Q^o_{pqrs..}$ (6)

For row 2, Table 1: $Q^n_{ijkl..} = (\pm)'(\pm)^{\Phi'} |a| a_{ip} a_{jq} a_{kr} a_{ls} Q^o_{pqrs..}$ (7)

For row 3, Table 1: $Q^n_{ijkl..} = (\pm)'(\pm)^{\Phi} a_{ip} a_{jq} a_{kr} a_{ls} Q^o_{pqrs..}$ (8)

For row 4, Table 1: $Q^n_{ijkl..} = a_{ip} a_{jq} a_{kr} a_{ls} Q^o_{pqrs..}$ (9)

**Supplementary Information** is linked to the online version of the paper at www.nature.com/nature

**Acknowledgements** The authors acknowledge financial support from the National Science Foundation through the MRSEC program DMR-0820404 and grant DMR-0908718. Discussions with Craig J. Fennie and A. M. Glazer and gratefully acknowledged.

**Author Contributions** V.G. conceived the idea of rotation reversal symmetry, the roto groups, and their influence on properties. D. B. L. critiqued and helped develop formal definitions for these concepts, and with the derivation of the symmetries using group theoretical methods. V.G. and D.B.L. wrote the Letter.

**Competing interests statement** The authors declare that they have no competing financial interests.

**Correspondence** and requests for materials should be addressed to V.G. (vgopalan@psu.edu).

## FIGURE LEGENDS

**Figure 1 | Rotation reversal symmetry and other antisymmetry operations**. When the colors of atoms in **a** denote the orientation of magnetic spins, +M (aqua, spin up, left handed loop) and –M (orange, spin down, right handed loop), the time reversal symmetry operation $1'$ will switch them to –M and +M, respectively, as in **b**. If the colors of a structural motif, as in **c**, instead denoted *static* rotations, +Φ and -Φ, the rotation reversal symmetry operation $1^{\Phi}$ switches them to -Φ and +Φ, respectively, as in **d**. In reversing polar and axial vectors (rows), either time dependent or static (columns), $1^{\Phi}$ is the missing symmetry operation (colored box) among the three required, the other two being inversion, $\bar{1}$ and time reversal, $1'$.

**Figure 2 | Roto groups and property classification**. **a** Based on the presence or absence of the three antisymmetry operations, $1^{\Phi}$, $\bar{1}$, and $1'$, eight types of groups are defined: triple grey (TG), magnetic (M), roto (R), magneto-roto (MR), polar (P), magneto-polar (MP), roto-polar (RP), and magneto-roto-polar (MRP). The number *x* of the MR groups is presently unknown. **b** The number of space groups that are symmetry groups of arrangements of spins, static rotations, electric dipoles, and combinations thereof, is given. In addition, the number of corresponding point groups (also see supplementary Table 1 for a listing) is given that are invariance groups of a spin, static rotation, electric dipole, and of combinations thereof.

**Figure 3 | Symmetries in antidistorted cubic perovskite lattices**. Antidistorted octahedra are shown disconnected for clarity. The Glazer rotation $a_o^+a_o^+c_o^+$ is depicted in **a,** where orange and aqua correspond to left and right handed rotations of octahedra, respectively, about the axes. Loops with arrows indicate the sense of rotation, and number of loops indicates the magnitude of

rotation. The exploded symmetries, **b,** reveal that while the conventional symmetry is $Immm1'$, the complete symmetry is $I4^{\Phi}/mmm^{\Phi}1'$. Panels **c,** show $a_o^+a_o^+c_o^+$ with magnetic spins inside each octahedron, with an exploded view of its symmetries in **d**. The conventional symmetry is $Im'm'm$, but the complete symmetry is $I4^{\Phi}/mm'm^{\Phi'}$.

**Figure 4 | Symmetries in helices and spirals**. A solenoid can possess **a,** static left-handed winding (orange), and carries a charge current (yellow arrows). The $1^{\Phi}$ only switches the static winding. The $1'$ symmetry only switches the charge current. The $1^{\Phi'}$ symmetry switches both. A single infinite helix, **b,** with a pitch of $\Lambda$, and a general infinite double helix, **c,** with arbitrary shift between the two helices in the $z$-direction, both have a point group of $(\infty/m^{\Phi})m^{\Phi}2$. A finite continuous helix with integral windings, **d,** has a point group of $m^{\Phi}m^{\Phi}2$. A finite planar spiral in **e**, has a point group of $mm^{\Phi}2^{\Phi}$. A single left-handed infinite helix, **f,** with $3_2$ screw axis with fractional atom positions at 0, 1/3, and 2/3 of the pitch is transformed by $1^{\Phi}$ operation to right handed $3_1$ as in **g**. The complete point group is $(\bar{6}^{\Phi}/m^{\Phi})m^{\Phi}2$.

**Figure 1**

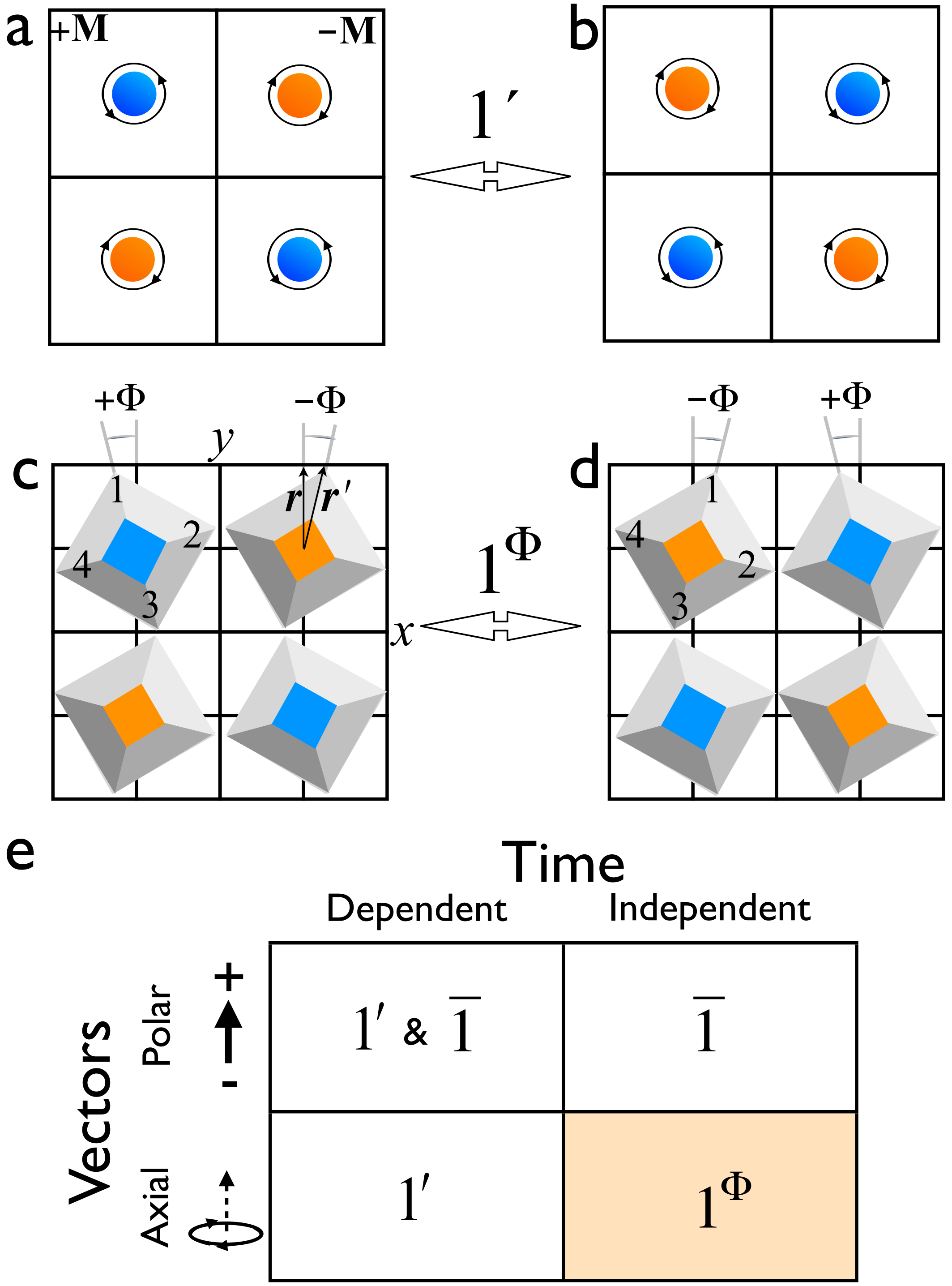

**Figure 2**

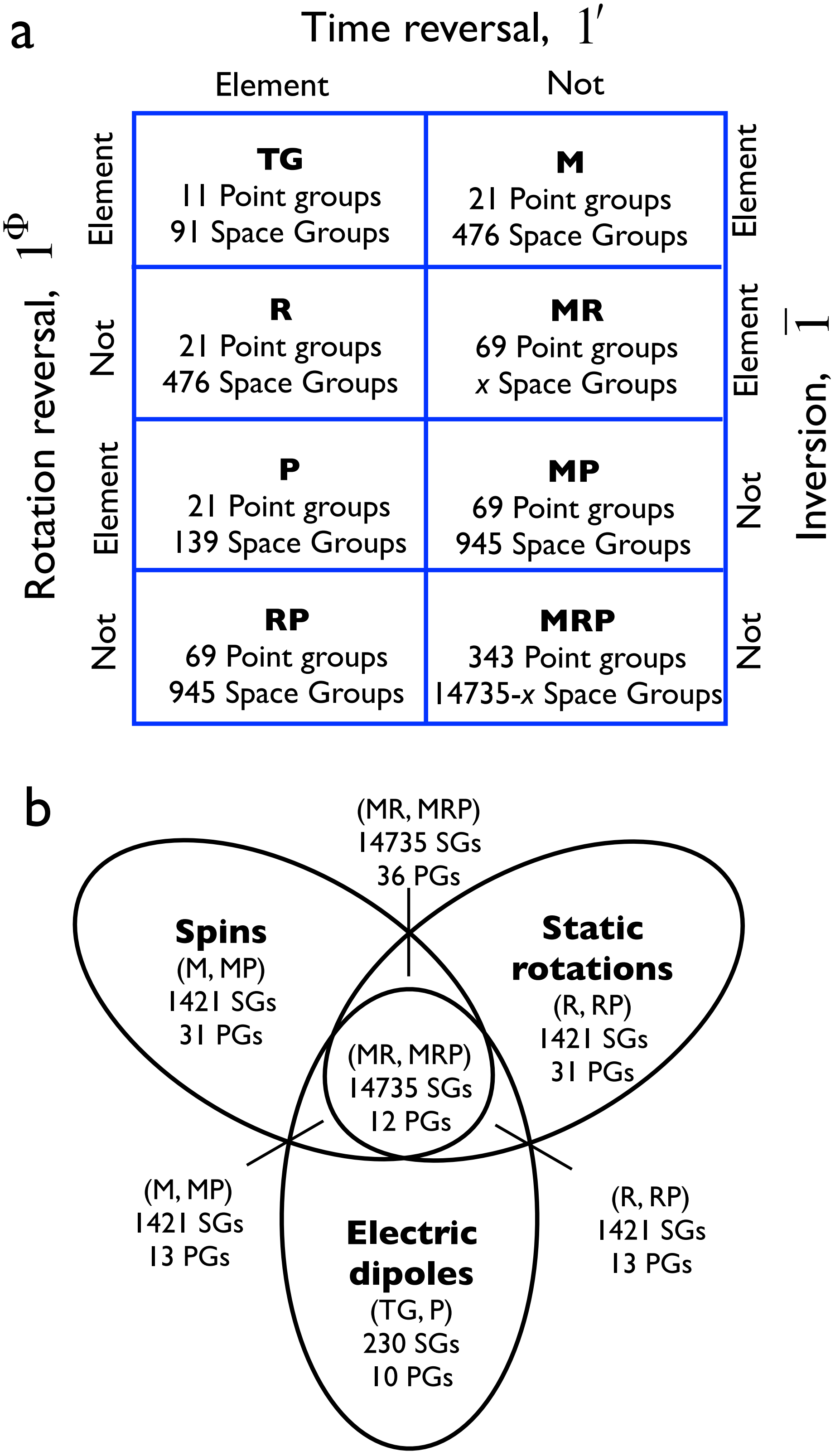

**Figure 3**

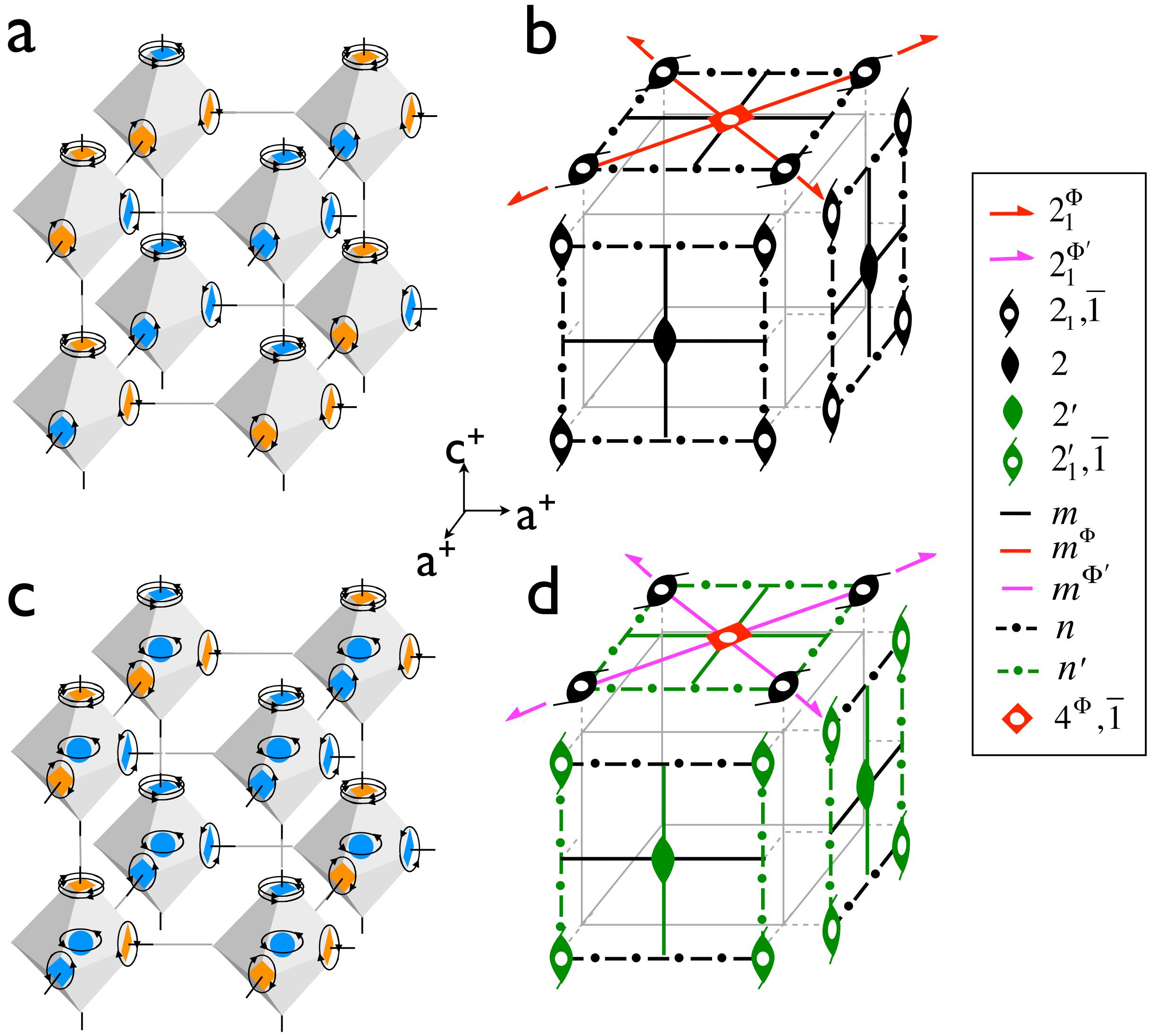

**Figure 4**

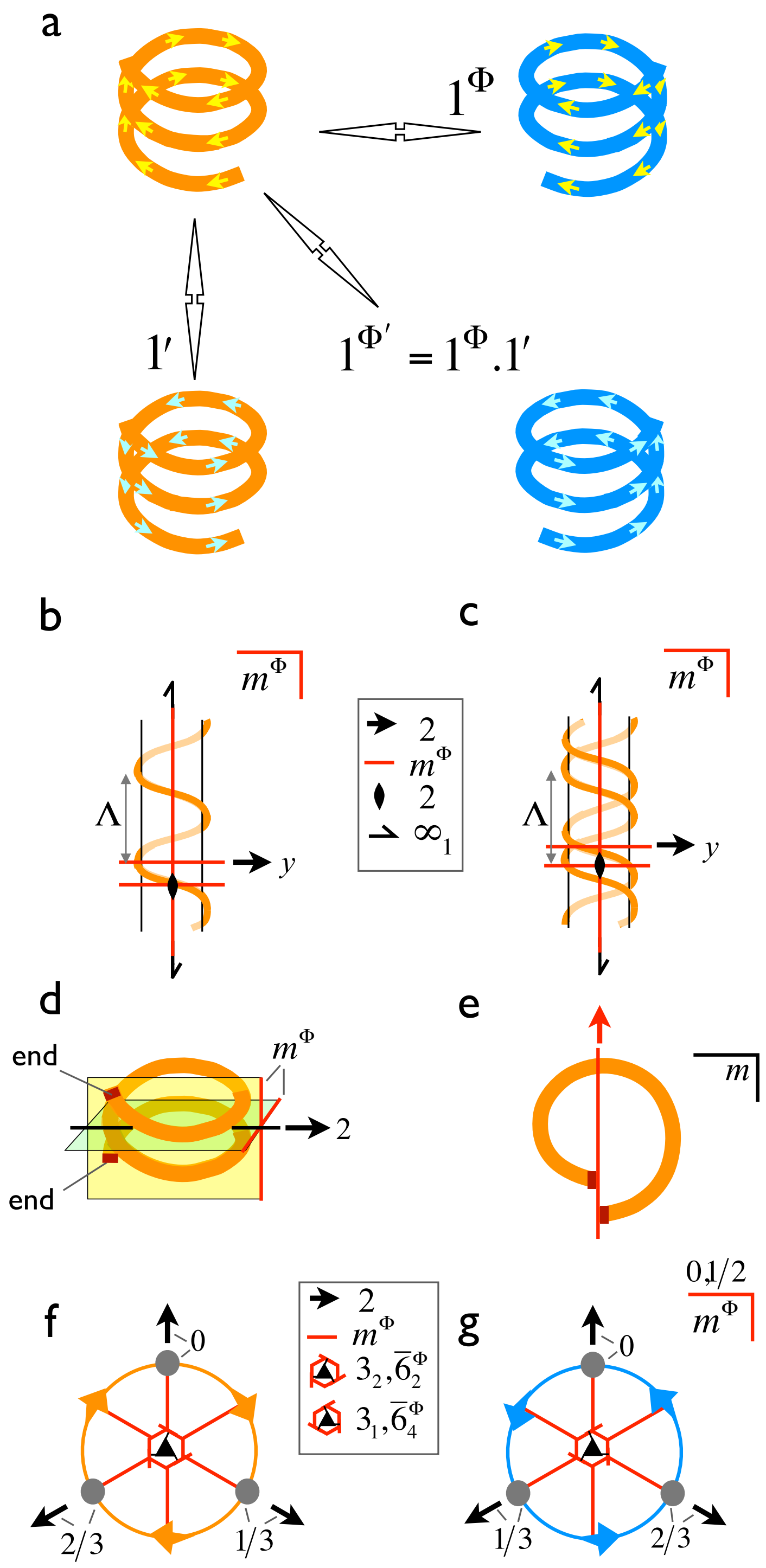

*Supplementary Information, Table of Contents*

**Supplementary Table 1**

**List of point groups of space groups indicated in Figure 1e of arrangements of spins (*S*), electric dipole (*P*), static rotations (Φ), and combinations thereof.**

| *S* only | Φ only | *P* only | *S*+*P* | P+Φ | Φ+*S* | Φ+*P*+*S* |
|---|---|---|---|---|---|---|
| $\bar{1}1^{\Phi}$ | $\bar{1}1'$ | $11'1^{\Phi}$ | $2m'm'1^{\Phi}$ | $2m^{\Phi}m^{\Phi}1'$ | $\bar{1}$ | $1$ |
| $2/m1^{\Phi}$ | $2/m1'$ | $21'1^{\Phi}$ | $4m'm'1^{\Phi}$ | $4m^{\Phi}m^{\Phi}1'$ | $2/m$ | $2$ |
| $2'/m'1^{\Phi}$ | $2^{\Phi}/m^{\Phi}1'$ | $m1'1^{\Phi}$ | $3m'1^{\Phi}$ | $3m^{\Phi}1'$ | $\bar{4}$ | $m$ |
| $mm'm'1^{\Phi}$ | $mm^{\Phi}m^{\Phi}1'$ | $31'1^{\Phi}$ | $6m'm'1^{\Phi}$ | $6m^{\Phi}m^{\Phi}1'$ | $4/m$ | $3$ |
| $\bar{4}1^{\Phi}$ | $\bar{4}1'$ | $41'1^{\Phi}$ | $11^{\Phi}$ | $11'$ | $\bar{6}$ | $4$ |
| $4/m1^{\Phi}$ | $4/m1'$ | $61'1^{\Phi}$ | $21^{\Phi}$ | $21'$ | $\bar{3}$ | $6$ |
| $2'm'1^{\Phi}$ | $2^{\Phi}m^{\Phi}1'$ | $2mm1'1^{\Phi}$ | $m1^{\Phi}$ | $m1'$ | $1$ | $2'$ |
| $4/mm'm'1^{\Phi}$ | $4/mm^{\Phi}m^{\Phi}1'$ | $4mm1'1^{\Phi}$ | $2'1^{\Phi}$ | $2^{\Phi}1'$ | $2$ | $m'$ |
| $\bar{6}1^{\Phi}$ | $\bar{6}1'$ | $3m1'1^{\Phi}$ | $mm'2'1^{\Phi}$ | $m^{\Phi}1'$ | $m$ | $2^{\Phi}$ |
| $\bar{3}1^{\Phi}$ | $\bar{3}1'$ | $6mm1'1^{\Phi}$ | $41^{\Phi}$ | $mm^{\Phi}2^{\Phi}1'$ | $4$ | $m^{\Phi}$ |
| $6/m1^{\Phi}$ | $6/m1'$ | | $31^{\Phi}$ | $41'$ | $3$ | $2^{\prime\Phi}$ |
| $\bar{3}m'1^{\Phi}$ | $m^{\Phi}1'$ | | $61^{\Phi}$ | $31'$ | $6$ | $m^{\prime\Phi}$ |
| $\bar{6}2'm'1^{\Phi}$ | $\bar{6}2^{\Phi}m^{\Phi}1'$ | | | $61'$ | $2mm$ | |
| $6/mm'm'1^{\Phi}$ | $6/mm^{\Phi}m^{\Phi}1'$ | | | | $4mm$ | |
| $2m'm'1^{\Phi}$ | $2m^{\Phi}m^{\Phi}1'$ | | | | $3m$ | |
| $4m'm'1^{\Phi}$ | $4m^{\Phi}m^{\Phi}1'$ | | | | $6mm$ | |
| $3m'1^{\Phi}$ | $3m^{\Phi}1'$ | | | | $2'$ | |
| $6m'm'1^{\Phi}$ | $6m^{\Phi}m^{\Phi}1'$ | | | | $m'$ | |
| $22'2'1^{\Phi}$ | $22^{\Phi}2^{\Phi}1'$ | | | | $mm'2'$ | |
| $42'2'1^{\Phi}$ | $42^{\Phi}2^{\Phi}1'$ | | | | $2^{\Phi}$ | |
| $32'1^{\Phi}$ | $32^{\Phi}1'$ | | | | $m^{\Phi}$ | |
| $62'2'1^{\Phi}$ | $62^{\Phi}2^{\Phi}1'$ | | | | $mm^{\Phi}2^{\Phi}$ | |
| $11^{\Phi}$ | $11'$ | | | | $2^{\prime\Phi}/m^{\prime\Phi}$ | |
| $21^{\Phi}$ | $21'$ | | | | $mm^{\prime\Phi}m^{\prime\Phi}$ | |
| $m1^{\Phi}$ | $m1'$ | | | | $\bar{4}2^{\prime\Phi}m^{\prime\Phi}$ | |
| $2'1^{\Phi}$ | $2^{\Phi}1'$ | | | | $4/mm^{\prime\Phi}m^{\prime\Phi}$ | |
| $m'1^{\Phi}$ | $m^{\Phi}1'$ | | | | $\bar{3}m^{\prime\Phi}$ | |
| $mm'2'1^{\Phi}$ | $mm^{\Phi}2^{\Phi}1'$ | | | | $\bar{6}2^{\prime\Phi}m^{\prime\Phi}$ | |
| $41^{\Phi}$ | $41'$ | | | | $6/mm^{\prime\Phi}m^{\prime\Phi}$ | |
| $31^{\Phi}$ | $31'$ | | | | $22^{\prime\Phi}2^{\prime\Phi}$ | |
| $61^{\Phi}$ | $61'$ | | | | $42^{\prime\Phi}2^{\prime\Phi}$ | |
| | | | | | $32^{\prime\Phi}$ | |
| | | | | | $62^{\prime\Phi}2^{\prime\Phi}$ | |
| | | | | | $2^{\prime\Phi}$ | |
| | | | | | $m^{\prime\Phi}$ | |
| | | | | | $mm^{\prime\Phi}2^{\prime\Phi}$ | |

**Supplementary Table 2.**

**Glazer antidistortions in centrosymmetric, non-magnetic cubic systems and their roto space groups (SG).**

| Glazer Tilt | Old SG | New Roto SG | Glazer Tilt | Old SG | New Roto SG |
|---|---|---|---|---|---|
| 1. $a_o^+b_o^+c_o^+$ | $Immm1'$ | $Immm1'$ | 13. $a_o^-b_o^-b_o^-$ | $C2/c1'$ | $C_p2^\Phi/m^\Phi1'$ |
| 2. $a_o^+a_o^+c_o^+$ | $Immm1'$ | $I4^\Phi/mmm^\Phi1'$ | 14. $a_o^-a_o^-a_o^-$ | $R\bar{3}c1'$ | $R_R\bar{3}m^\Phi1'$ |
| 3. $a_o^+a_o^+a_o^+$ | $Im\bar{3}1'$ | $Im\bar{3}m^\Phi1'$ | 15. $a_o^ob_o^+c_o^+$ | $Immm1'$ | $C_Immm1'$ |
| 4. $a_o^+b_o^+c_o^-$ | $Pmmn1'$ | $P_{2c}mm21'$ | 16. $a_o^ob_o^+b_o^+$ | $I4/mmm1'$ | $P_I4/mmm1'$ |
| 5. $a_o^+a_o^+c_o^-$ | $P4_2/nmc1'$ | $P_{2c}4^\Phi/nmm^\Phi1'$ | 17. $a_o^ob_o^+c_o^-$ | $Cmcm1'$ | $P_Amm21'$ |
| 6. $a_o^+b_o^+b_o^-$ | $Pmmn1'$ | $P_{2c}mm21'$ | 18. $a_o^ob_o^+b_o^-$ | $Cmcm1'$ | $P_Amm21'$ |
| 7. $a_o^+a_o^+a_o^-$ | $P4_2/nmc1'$ | $P_{2c}4^\Phi/nmm^\Phi1'$ | 19. $a_o^ob_o^-c_o^-$ | $C2/m1'$ | $P_C2/m1'$ |
| 8. $a_o^+b_o^-c_o^-$ | $P2_1/m1'$ | $P_{2a}2_1/m1'$ | 20. $a_o^ob_o^-b_o^-$ | $Imma1'$ | $C_Im^\Phi mm1'$ |
| 9. $a_o^+a_o^-c_o^-$ | $P2_1/m1'$ | $P_{2a}2_1/m1'$ | 21. $a_o^oa_o^oc_o^+$ | $P4/mbm1'$ | $P_P4^\Phi/mmm^\Phi1'$ |
| 10. $a_o^+b_o^-b_o^-$ | $Pnma1'$ | $C_pm^\Phi c^\Phi m1'$ | 22. $a_o^oa_o^oc_o^-$ | $I4/mcm1'$ | $P_I4/mm^\Phi m^\Phi1'$ |
| 11. $a_o^+a_o^-a_o^-$ | $Pnma1'$ | $C_pm^\Phi c^\Phi m1'$ | 23. $a_o^oa_o^oa_o^o$ | $Pm\bar{3}m1'$ | $Pm\bar{3}m1'1^\Phi$ |
| 12. $a_o^-b_o^-c_o^-$ | $P\bar{1}1'$ | $P_{2s}\bar{1}1'$ | | | |

**Table 2** | Conventional and new space group assignments for the 23 non-magnetic Glazer antidistortive tilt rotation systems[5,28]. Each Glazer system marked in yellow is one of a pair of systems, e.g. #17 and #18, which have identical conventional and new space group assignments [29]. Those systems marked in blue all have the same conventional space group assignment and distinct new space group assignments.

**Supplementary Table 3**. **“Roto” property tensors and their symmetry.**

Rank of property tensors

| Properties destroyed by | First | Second | Third | Fourth |
|---|---|---|---|---|
| $1^{\Phi}$ & $1^{\Phi'}$ | Static Rotations<br>$\Phi_i$ | linear Rotoelectricity<br>$P_i = Q_{ij}\Phi_j$ | Piezorotation<br>$\Phi_i = Q_{ijk}\sigma_{jk}$ | Piezorotoelectricity<br>$P_i = Q_{ijkl}\sigma_{jk}\Phi_l$ |
| $1'$ & $1^{\Phi'}$ | Double Curl of $\boldsymbol{M}$<br>$Q_i = (\nabla \times \nabla \times \boldsymbol{M})_i$ | Rototorroidal-magnetism<br>$(\nabla \times \boldsymbol{M})_i = Q_{ij}\Phi_j$ | quadratic Rotomagnetism<br>$M_i = Q_{ijk}\Phi_j\Phi_k$ | quadratic Rotomagnetoelectricity<br>$P_i = Q_{ijkl}\Phi_j\Phi_k M_l$ |
| $1^{\Phi}$ & $1'$ | Torroidal magnetism<br>$Q_i = (\nabla \times \boldsymbol{M})_i$ | linear Rotomagnetism<br>$M_i = Q_{ij}\Phi_j$ | linear Roto-magnetoelectricity<br>$P_i = Q_{ijk}\Phi_j M_k$ | Piezorotomagnetism<br>$M_i = Q_{ijkl}\sigma_{jk}\Phi_l$ |
| None | Inverse DM interaction<br>$Q_i = (\boldsymbol{r}_{12} \times \boldsymbol{M}_1 \times \boldsymbol{M}_2)_i$ | Rotomagnetic canting<br>$(\boldsymbol{L} \times \boldsymbol{M})_i = Q_{ij}\Phi_j$ | quadratic Rotoelectricity<br>$P_i = Q_{ijk}\Phi_j\Phi_k$ | Rotostriction<br>$\varepsilon_{ij} = Q_{ijkl}\Phi_k\Phi_l$ |

**Table 3** I Each “roto” property tensor, $Q$ is named, and then defined below it. Variables are defined as follows: $\Phi$ - static rotation, $P$ – polarization, $M$ – magnetization, $\boldsymbol{r}_{12}$ – distance vector between magnetic moments $M_1$ at location 1 and $M_2$ at location 2, $\boldsymbol{L} = \boldsymbol{M}_1 - \boldsymbol{M}_2$, $\boldsymbol{M} = \boldsymbol{M}_1 + \boldsymbol{M}_2$, $\sigma$ – stress, $\varepsilon$ – strain. Subscripts $i, j, k, l$ indicate the component, and can each be one of the three orthogonal coordinates $x$, $y$, or $z$.

**Supplementary Discussion 1.**

**Roto- group assignments for normal modes of $NaNbO_3$ and $BiFeO_3$**

The room temperature structure of $NaNbO_3$ is *Pbcm*.[30] Normal mode analysis[30] shows that this arises from the symmetry intersection of the two primary octahedral antidistortion rotations: R4+ (*Imma*) and T4 (*I4/mcm*). We note that R4+ mode is an $a_o^- a_o^- c_o^o$ type Glazer distortion which doubles all the lattice parameters, and the T4 mode has an alternating sequence of $a_o^o a_o^o c_o^-$ and $a_o^o a_o^o c_o^+$ distortions, which quadruples the lattice parameter along the *c*-axis after distortion. The intersection of the symmetries of these primary distortions with the largest amplitudes determines the *true* observed symmetry of the distorted phase, and the secondary distortions with weak amplitudes typically have higher symmetry and are consistent with the symmetry of the primary distortion. The superspace groups approach[31] yields results consistent with the normal mode analysis.[30] We determine their new space group symmetries to be R4+ ( $C_I m^{\Phi} mm1'$) and T4 ($P_I 4/mm^{\Phi} m^{\Phi} 1'$). The intersection of these two symmetries gives a new symmetry group for $NaNbO_3$ of $C_P m^{\Phi} cm1'$ (R), which has twice the number of symmetry elements as the conventional group $Pbcm1'$ assigned to this structure.

$BiFeO_3$ with $a_+^- a_+^- a_+^-$ octahedral rotations and ignoring spins has a conventional symmetry of $R3c1'$; considering both the spins and the antidistortion, the conventional magnetic symmetry is $Cc$ or $Cc'$ for spins in or perpendicular to the *c*-glide plane, respectively[32] when canting is allowed. We derive an RP-group symmetry of $R_R 3m^{\Phi} 1'$ for the octahedral rotations, ignoring spins. The MP symmetry (ignoring rotations, but allowing canting) remains $Cc$ or $Cc'$ as before. The composite MRP-group symmetry (RP∩MP) is $Cc$ or $Cc'$.

**Supplementary Discussion 2.**

**Defined roto properties in Supplementary Table 3 and literature measurements in antidistorted perovskites**

The literature examples cited below are limited to antidistorted structures. Among the "roto" properties that are invariant under $1'$, $1^{\Phi}$ and $1^{\Phi'}$ (4th row of supplementary Table 3), Haun[20] measured the coefficients of *rotostriction tensor*, $Q_{ijkl}$ in the $Pb(Zr_xTi_{1-x})O_3$ (PZT) ferroelectric system, arising from the energy term $Q_{ijkl}\sigma_{ij}\Phi_k\Phi_l$ . Haun[20] also measured the *biquadratic rotoelectric tensor* $\gamma_{ijkl}$ (without naming it), arising from the energy interaction term $\gamma_{ijkl}P_iP_j\Phi_k\Phi_l$. The *quadratic rotoelectric tensor* $Q_{jkl} = \gamma_{ijkl}P_i^o$ in supplementary Table 3 also arises from this interaction in a polar state with a spontaneous polarization, $P_i^o$. In nanoscale multilayer $PbTiO_3/SrTiO_3$ films, the quadratic rotoelectric effect arising from the energy interaction, $-Q_{ijk}P_i\Phi_j\Phi_k$ was shown by Bousquet et. al. to lead to improper ferroelectricity through octahedral rotations[21]. The *rotomagnetoelectric effect*, arising from the energy term $-Q_{ijk}P_i\Phi_j(\boldsymbol{L}\times\boldsymbol{M})_k$ gives rise to a coupled ferroelectric polarization, $P_i$, antiferromagnetic vector, $\boldsymbol{L}$ and weak magnetism through spin canting, $\boldsymbol{M}$ in multiferoic $FeTiO_3$[22, 33]. The weak ferromagnetism in $BiFeO_3$ arising from the *rotomagnetic canting* of the antiferromagnetic spins mediated by octahedral rotations $\Phi_i$ through the Dyaloshinskii-Moriya (DM) energy interaction, $-Q_{ij}\Phi_i(\boldsymbol{L}\times\boldsymbol{M})_j$[32]. *Quadratic pyrorotation tensor* $Q_{ij}$, given by $\Phi_i\Phi_j = Q_{ij}(T_c - T)$, is seen in materials below a rotational phase transition temperature, $T_c$ such as in titanates[34,35,36,37,38] and aluminates[38]. Similarly, *quadratic hydrostatic rotation*, i.e. rotation under pressure, $p$, defined by $\Phi_i\Phi_j = Q_{ij}(p - p_c)$, has been widely reported[39,40,41,42]. An *inverse DM interaction* of the type $P_i \propto (\boldsymbol{r}_{12}\times\boldsymbol{S}_1\times\boldsymbol{S}_2)_i$ between spins at locations 1 and 2, separated by a vector $\boldsymbol{r}_{12}$ is observed to give rise to a polarization $P_i$ in spin cycloidal ferroelectrics such as $TbMnO_3$, $DyMnO_3$, $Ni_3V_2O_8$, $CuFeO_2$, and $Ba_{0.5}Sr_{1.5}Zn_2Fe_{12}O_{22}$[43]. *Biquadratic rotomagnetic* interaction ($Q_{ijkl}\Phi_i\Phi_j M_k M_l$) is present in all magnetic materials with rotations.

The row 1 of Supplementary Table 3 lists properties that invert under $1^{\Phi}$ and $1^{\Phi'}$ but are invariant under $1'$, such as static octahedral rotations, $\Phi_i$, in perovskites. The *linear rotoelectric effect*, $Q_{jkl} = \gamma_{ijkl} P_i^o \Phi_k^o$ arises from the quadratic rotoelectric effect biased by a spontaneous polarization, $P_i^o$ and a rotational distortion, $\Phi_i^o$. Both linear and quadratic rotoelectric effect contributions to a polarization anomaly at the antidistortive phase transition has been observed in ferroelectric $Pb_{1-y}(Zr_{1-x}Ti_x)_{1-y}Nb_yO_3$[34,44]. A combination of linear and third order rotoelectric effects has also been shown as the source of improper ferroelectricity in $YMnO_3$[23]. *Linear hydrostatic rotation*, defined as $\Phi_i = Q_i(p - p_c)$ is also reported above a critical pressure $p>p_c$[45,46,47,48]. *Piezorotation* has been predicted in $LaAlO_3$ under biaxial strain[24]. The magnitude and sign of superexchange interaction, $-J_{ij}S_iS_j$, between Mn atom spins in $AMnF_4$ (A-alkali atoms) is dependent on the Mn-F-Mn tilt angles, which in turn is related to the octahedral rotations, $\Phi$, as $J_{ij} = Q_{ijk}(\Phi_k - \Phi_k^o)$, which can be termed *rotomagnetic exchange*[49]. *Chiral optical sum frequency generation* (SFG) is proportional to $\mathbf{p}_1 \cdot (\mathbf{p}_2 \times \mathbf{p}_3)$, where $\mathbf{p}_{1,2,3}$ are three different electric dipole transition moments generated by the light fields[25]. The cross product term inverts under $1^{\Phi}$ and $1^{\Phi'}$, hence its sensitivity to chirality.

The second row in supplementary Table 3 includes properties that invert under $1'$ and $1^{\Phi'}$ but are invariant under $1^{\Phi}$. Magnetoelectricity, piezomagnetism, and pyromagnetism are well known[10]. *Optical activity* of handed crystals is defined by the rotary power $R = \mathrm{Im}\{\mathbf{p} \cdot \mathbf{m}\}$, where $\mathbf{p}$ and $\mathbf{m}$ are electric and magnetic dipole transition moments generated by light field[25]. Double curl of magnetization occurs in Maxwell's electromagnetic wave equation in matter[50]. *Quadratic rotomagnetism* is a manifestation of *biquadratic rotomagnetism* biased by a spontaneous magnetization, $M_i^o$ (energy term, $-Q_{ijkl}M_i^o B_j \Phi_k \Phi_l$ ) in a magnetically ordered phase with rotational distortions. *Quadratic rotomagnetoelectricity* can arise in antiferromagnetic magnetoelectrics such as $FeTiO_3$ and $BiFeO_3$ from the energy term $-Q_{ijklm} E_i \Phi_j \Phi_k M_l L_m$[22,33].

The third row in Supplementary Table 3 are properties that invert under both $1'$ and $1^{\Phi}$, but are invariant under $1^{\Phi'}$. *Current density* (current per unit area), is defined by Ampere's law as $J_i = (\nabla \times B)_i / \mu_o$, where $\mu_o$ is the magnetic permeability of free space[50]. Note that both the cross product and the area are axial vectors that invert under $1^{\Phi}$. *Torroidal magnetic moment* was reported in $LiCoPO_4$[26] $BiFeO_3$ and $FeTiO_3$[51] and others. *Linear rotomagnetism* is a manifestation of biquadratic rotomagnetism biased by a spontaneous magnetization, $M_i^o$ and spontaneous rotation, $\Phi_i^o$ . For example, it is observed in $Se_{1-x}$ $Te_xCuO_3$, where as octahedral rotations increase from $x$=1 to $x$=0, the saturation ferromagnetism increases linearly[27]. *Linear rotomagnetoelectricity* can arise from the energy term $-Q_{ijklm} E_i \Phi_j \Phi_k^o M_l L_m^o$.

**Supplementary Discussion 3**

**Space group analysis of tensor properties**

This section derives the predicted tensor forms for quadratic electrorotation ($\Phi_i = Q_{ijk} E_j E_k$). Let us consider the example of Glazer rotations $a_o^+ a_o^+ c_o^+$ in a cubic pervoskite structure in Fig. 3a. The conventional space group assignment is the orthorhombic space group $Immm1'$, while our new roto group assignment is a tetragonal space group $I4^{\Phi}/mmm^{\Phi}1'$. The centre of mass of the octahedral are at the Wyckoff location *k*, which has 8 equivalent positions[52]. These eight octahedral positions in roto space group and $I4^{\Phi}/mmm^{\Phi}1'$ in $Immm1'$:

**1**: (¼, ¼, ¼), **2**: (¾, ¾, ¼), **3**: (¾, ¼, ¾), **4**: (¼, ¾, ¾), **5**: (¾,¾,¾), **6**: (¼,¼,¾), **7**: (¼,¾,¼), **8**: (¾,¼,¼)

**Supplementary Equation 1**

The static rotations of the octahedra in Fig. 3a lead to the following local order parameters:

$$\Phi_x = (\Phi_{1,x} + \Phi_{4,x} + \Phi_{5,x} + \Phi_{8,x}) - (\Phi_{2,x} + \Phi_{3,x} + \Phi_{6,x} + \Phi_{7,x})$$
$$\Phi_y = (\Phi_{1,y} + \Phi_{3,y} + \Phi_{5,y} + \Phi_{7,y}) - (\Phi_{2,y} + \Phi_{4,y} + \Phi_{6,y} + \Phi_{8,y})$$
$$\Phi_z = (\Phi_{1,z} + \Phi_{2,z} + \Phi_{5,z} + \Phi_{6,z}) - (\Phi_{3,z} + \Phi_{4,z} + \Phi_{7,z} + \Phi_{8,z})$$

**Supplementary Equation 2**

Here the subscripts indicate the octahedral number and the component axis. The following supplementary table 4 shows how these order parameters transform when operated upon by the (000)+ symmetry operations of space group $Immm1'$.

**Supplementary Table 4**: Transformations of the order parameter components in supplementary Eq. (2), and electric field components $E_i$ ($i$=$x,y,z$) under the (000)+ set of operations of the space group $Immm1'$.

| | 1 | $2_z$ | $2_y$ | $2_x$ | $\bar{1}$ | $m_z$ | $m_y$ | $m_x$ |
|---|---|---|---|---|---|---|---|---|
| $\Phi_z$ | $\Phi_z$ | $\Phi_z$ | $-\Phi_z$ | $-\Phi_z$ | $\Phi_z$ | $\Phi_z$ | $-\Phi_z$ | $-\Phi_z$ |
| $\Phi_x$ | $\Phi_x$ | $-\Phi_x$ | $-\Phi_x$ | $\Phi_x$ | $\Phi_x$ | $-\Phi_x$ | $-\Phi_x$ | $\Phi_x$ |
| $\Phi_y$ | $\Phi_y$ | $-\Phi_y$ | $\Phi_y$ | $-\Phi_y$ | $\Phi_y$ | $-\Phi_y$ | $\Phi_y$ | $-\Phi_y$ |
| $E_x$ | $E_x$ | $-E_x$ | $-E_x$ | $E_x$ | $-E_x$ | $E_x$ | $E_x$ | $-E_x$ |
| $E_y$ | $E_y$ | $-E_y$ | $E_y$ | $-E_y$ | $-E_y$ | $E_y$ | $-E_y$ | $E_y$ |
| $E_z$ | $E_z$ | $E_z$ | $-E_z$ | $-E_z$ | $-E_z$ | $-E_z$ | $E_z$ | $E_z$ |

**Supplementary Table 5**. Transformations of the order parameter components in supplementary Eq. (2), and electric field components $E_i$ ($i$=$x,y,z$) under the (000)+ set of operations of the space group $I4^{\Phi}/mmm^{\Phi}1'$ that are not present in $Immm1'$.

| | $4_z^{\Phi}$ | $4_z^{-1\Phi}$ | $2_{xy}^{\Phi}$ | $2_{\bar{x}y}^{\Phi}$ | $\bar{4}_z^{\Phi}$ | $\bar{4}_z^{-1\Phi}$ | $m_{xy}^{\Phi}$ | $m_{\bar{x}y}^{\Phi}$ |
|---|---|---|---|---|---|---|---|---|
| $\Phi_z$ | $\Phi_z$ | $\Phi_z$ | $-\Phi_z$ | $-\Phi_z$ | $\Phi_z$ | $\Phi_z$ | $-\Phi_z$ | $-\Phi_z$ |
| $\Phi_x$ | $\Phi_y$ | $-\Phi_y$ | $\Phi_y$ | $-\Phi_y$ | $\Phi_y$ | $-\Phi_y$ | $-\Phi_y$ | $+\Phi_y$ |
| $\Phi_y$ | $-\Phi_x$ | $\Phi_x$ | $\Phi_x$ | $-\Phi_x$ | $-\Phi_x$ | $\Phi_x$ | $-\Phi_x$ | $\Phi_x$ |
| $E_x$ | $E_y$ | $-E_y$ | $E_y$ | $-E_y$ | $-E_y$ | $E_y$ | $E_y$ | $-E_y$ |
| $E_y$ | $-E_x$ | $E_x$ | $E_x$ | $-E_x$ | $E_x$ | $-E_x$ | $E_x$ | $-E_x$ |
| $E_z$ | $E_z$ | $E_z$ | $-E_z$ | $-E_z$ | $-E_z$ | $-E_z$ | $E_z$ | $E_z$ |

The space group $I4^{\Phi}/mmm^{\Phi}1'$ has more symmetries than $Immm1'$ whose transformations are listed in Supplementary Table 5 for the (000)+ set: The (½, ½, ½)+ set of operations for both the space groups duplicate the above tables and hence are not shown.

From the above tables, one can find invariant terms for each space group. For the space group $Immm1'$, the only invariant product terms under all operations of the space group are $\Phi_x E_y E_z$, $\Phi_x E_y E_z$, and $\Phi_y E_z E_x$. This can be checked by multiplying the corresponding rows in the Table 4 to find no change in the term for each symmetry operation. Similar analysis can be performed for the $I4^{\Phi}/mmm^{\Phi}1'$ space group. Therefore, the quadratic electrostriction tensor has the forms:

$$(Q_{ijk}) = \begin{pmatrix} 0 & 0 & 0 & Q_{14} & 0 & 0 \\ 0 & 0 & 0 & 0 & Q_{25} & 0 \\ 0 & 0 & 0 & 0 & 0 & Q_{36} \end{pmatrix}$$ for the $Immm1'$ space group.

$$(Q_{ijk}) = \begin{pmatrix} 0 & 0 & 0 & Q_{14} & 0 & 0 \\ 0 & 0 & 0 & 0 & -Q_{14} & 0 \\ 0 & 0 & 0 & 0 & 0 & 0 \end{pmatrix}$$ for the $I4^{\Phi}/mmm^{\Phi}1'$ group

**Supplementary Equation 3**

Clearly, the larger number of symmetry operations in $I4^{\Phi}/mmm^{\Phi}1'$ has reduced the number of tensor coefficients as well as given rise to new equalities between coefficients.

**Supplementary Discussion 4**

**Rotation Reversal Symmetry Operation on a Continuous Helix or a Spiral**

The static handedness of the helix can be represented by a following function:

$$\boldsymbol{r}(\Phi) = a\cos\Phi\hat{\boldsymbol{x}} + a\sin\Phi\hat{\boldsymbol{y}} + b|\Phi|\hat{\boldsymbol{z}}$$ **Supplementary Equation 4**

Here $a$ is the radius of the helix, and $b = \Lambda/2\pi$ is related to the pitch $\Lambda$ of the helix. The corresponding function $\boldsymbol{r}(\Phi)$ for a spiral is given as:

$$\boldsymbol{r}(\Phi) = a|\Phi|\cos\Phi\hat{\boldsymbol{x}} + a|\Phi|\sin\Phi\hat{\boldsymbol{y}}$$

**Supplementary Equation 5**

When $1^{\Phi}$ operates on these functions, the function transforms as follows: $\boldsymbol{r}(\Phi) \rightarrow \boldsymbol{r}(-\Phi)$, and in the process transforms a right-handed helix or spiral into a left-handed helix or a spiral.

**Supplementary Discussion 5.**

**Rotation reversal operation on a helical screw axis**

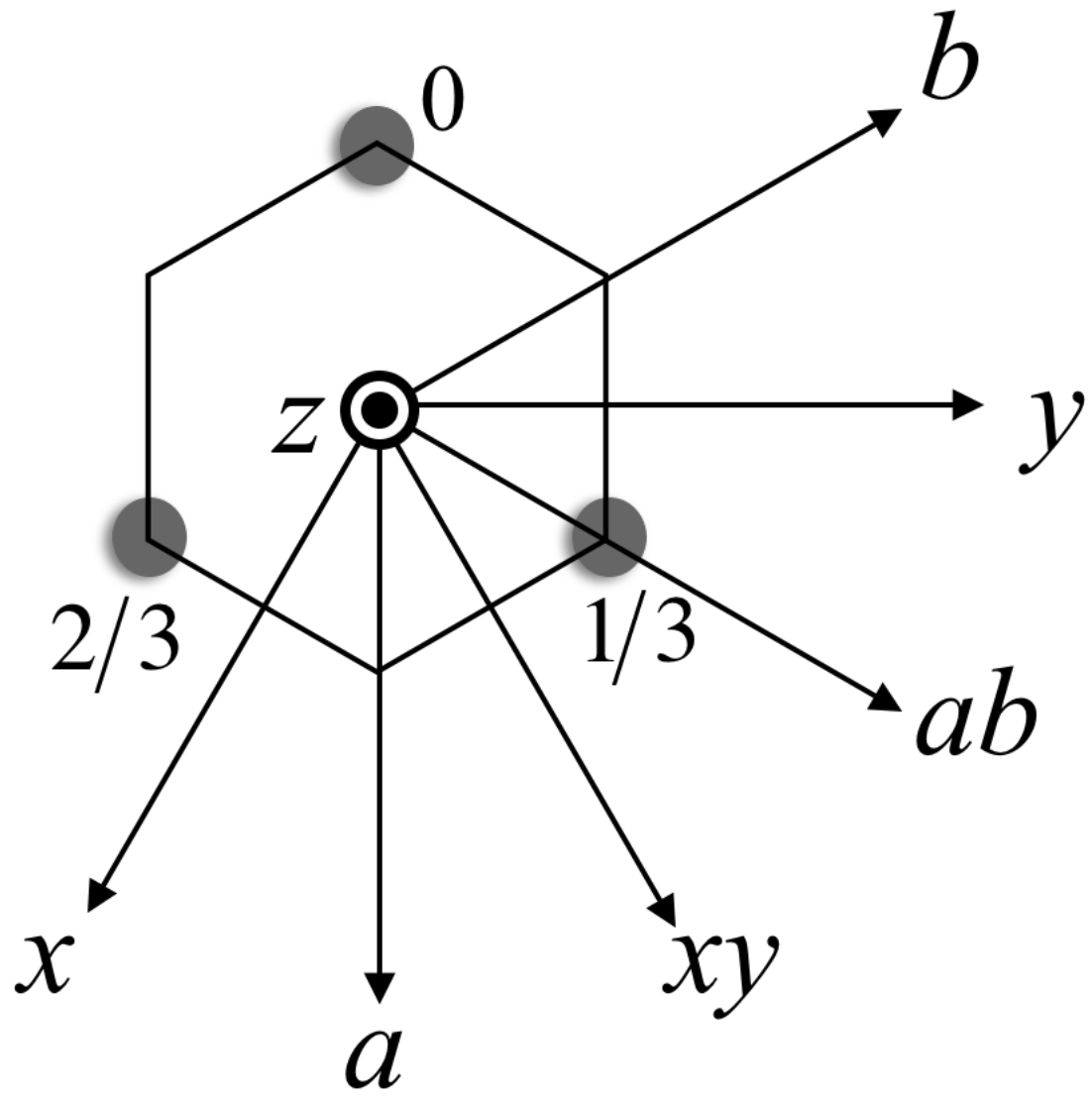

**Supplementary Figure 1** | A single infinite $3_2$ helix with atoms (blue circles) at 0, 1/3 and 2/3 fractional positions of the pitch along the *z*-axis. Various directions are labeled.

Consider a single infinite left handed helix (Fig. 4f and supplementary Fig. 1) with atoms at fractional positions 0, 1/3 and 2/3 of the pitch length along the screw axis, *z*. We define positive rotations to be counter-clockwise, the axes ($x, y, z$) to be right handed. In a symbol $(R|t)$, the rotation (or rotation inversion) $R$ acts first and then the translation $t$. Symmetry elements of this helix are the translations $(1|0\ 0\ n)$, where $n$ is an arbitrary integer and :

$$(1|0\ 0\ 0),\quad (3_z|0\ 0\ 2/3),\quad (3^2{}_z|0\ 0\ 1/3),\quad (2_a|0\ 0\ 0),\quad (2_{ab}|0\ 0\ 1/3),\quad (2_b|0\ 0\ 2/3)$$

**Supplementary equation 6**

These conventional elements generate the rod group $p3_212$ whose point group is 32.

However, we indentify additional symmetries. First we define the operation of $1^{\Phi}$ on the helix as follows: the "0" atom stays in its place, and from there instead of spiraling up as a left-handed helix (Fig. 4f) it spirals up as a right-handed helix (Fig. 4g), or vice versa. Note that with this definition, in general, $(R\,|\,T)1^{\Phi} \neq 1^{\Phi}(R\,|\,T)$. The additional symmetries are:

$(\bar{6}|0\ 0\ 1/3)1^{\Phi}$, $(\bar{6}^{-1}|0\ 0\ 2/3)1^{\Phi}$, $(m_z|0\ 0\ 0)1^{\Phi}$, $(m_x|0\ 0\ 2/3)1^{\Phi}$, $(m_{xy}|0\ 0\ 1/3)1^{\Phi}$, $(m_y|0\ 0\ 0)1^{\Phi}$

$1^{\Phi}(\bar{6}|0\ 0\ 2/3)$, $1^{\Phi}(\bar{6}^{-1}|0\ 0\ 1/3)$, $1^{\Phi}(m_z|0\ 0\ 0)$, $1^{\Phi}(m_x|0\ 0\ 1/3)$, $1^{\Phi}(m_{xy}|0\ 0\ 2/3)$, $1^{\Phi}(m_y|0\ 0\ 0)$

**Supplementary equation 7**

The translations $(1|0\ 0\ n)$ and the symmetry elements in Supplementary Equations 6 and 7 generate a group which we shall denote as $p\bar{6}_2^{\Phi}/m^{\Phi}m^{\Phi}2$ [53]. For point group elements, $R\cdot 1^{\Phi} = 1^{\Phi}\cdot R$, and consequently, the point group of this helical symmetry group is $\bar{6}^{\Phi}/m^{\Phi}m^{\Phi}2$. The order, number of elements, of this point group is twice that of the conventional point group 32, i.e. twelve instead of six.

We note that for a single helix, mirror operation such as $(m_y|0\ 0\ 0)$ performs the same operation as $1^{\Phi}$. Is there a need for the latter operation? The answer is *yes* for the same reason that though a mirror can also invert a single spin, time reversal 1′ is still required in order to flip spins in a lattice without translating them. Similarly, a single (finite or infinite) helix can be considered similar to a single spin in this respect. If a crystal is composed of many helices (such as protein crystals), then an appropriate mirror will reverse the handedness of all the helices, but also translate then space. In contrast, $1^{\Phi}$ will reverse the handedness of all the helices in a crystal without translating their center of mass in space.

**Supplementary Discussion 6.**

**Roto symmetries in continuous single and double helix structures**

**Single continuous infinite helix (Figure 4b)**

The conventional symmetry group of this helix is generated by the translation $(1|0\ 0\ \Lambda)$, and the two operations

1) $(\infty_z|0\ 0\ 1/\infty) = \lim_{N\to\infty}(N_z\,|0\ 0\ \ \Lambda/N)$ where $N_z$ is a rotation of $2\pi/N$ about the $z$-axis, and
2) $(2_y|0\ 0\ 0)$ a rotation of 180° about an axis perpendicular to the $z$-axis which intersects with the helix.

The point group is $\infty 2$, a group consisting of all possible rotations about the $z$-axis and all 2-fold rotations perpendicular to the $z$-axis. An additional generator of symmetries is the operation $(m_x|0\ 0\ 0)1^{\Phi}$. The point group of the new symmetry group of the helix is $\infty/m^{\Phi}m^{\Phi}2$ which has twice the number of operations as the conventional point group $\infty 2$. Note that $\bar{1}^{\Phi}$ is a symmetry element in the new point group.

**Double helix, continuous, infinite (Supplementary Figures 4c)**

For a general infinite double helix with an arbitrary shift of $z = f\Lambda\,(0<f<1)$ in the $z$-direction between the two identical helical strands, Fig. 4c, the symmetry is the same as the single strand helix as can be seen by using a coordinate system where the $y$-axis splits the two strands. A special case of the double helix is an infinite double-helix with a relative shift (along the $z$-axis) of half a pitch ($f$=0.5) between the two identical helices. The generators of the symmetry group of this helix are the same as the two previous helixes except that the translation generator $(1|0\ 0\ \Lambda)$ is replaced with $(1|0\ 0\ \Lambda/2)$. The point group thus remains the same, $\infty/m^{\Phi}m^{\Phi}2$.

**Crystalline polynucleotide fibers such as DNA**

Finally, we note that a crystalline polynucleotides fiber of deoxyribonucleic acid (DNA) is composed of double helices. However, each DNA helix is composed of a sequence of four chemical groups, adenine(A), thymine(T), guanine(G), and cytosine(C), with an integral or non-integral repeats per cycle of the helix [54]. The adenine on one of the helices bonds only with the thymine on the other helix, and similarly guanine bonds only with cytosine. As a consequence, the symmetry group will contain no horizontal mirror symmetries, nor 2-fold perpendicular symmetries. However, in an infinite strand of DNA with screw symmetry of $R_1$, (where $R$ can be integral or non-integral rotation axis), the infinite number of vertical mirror symmetries will remain. The point group is then $Rm^{\Phi}$.

**Supplementary Note 1.**

**List of articles referenced in supplementary discussions and figures**